\documentclass[12pt, twoside]{article}

\usepackage{amsfonts}
\newcommand{\bb}{\begin {minipage} {3cm}\begin{center}}
\newcommand{\ee}{\end{center}\end{minipage}}
\newcommand{\bc}{\begin {minipage} {2.5cm}\begin{center}}
\newcommand{\bd}{\end{center}\end{minipage}}

\newcommand {\D} {{\cal D}}

\newcommand {\q}{\begin{quote} \small}

\newcommand {\be}{\begin{equation}}
\newcommand {\e}{\end{equation}}
\newcommand {\bea}{\begin{eqnarray}}
\newcommand {\ea}{\end{eqnarray}}
\newcommand {\tr}{{\rm Tr}}

\newcommand {\g}{{\mathfrak g}}

\newcommand {\fract}[2]{\mbox{${\textstyle{\frac{#1}{#2}}}$}}

\begin{document}
\newtheorem {lemma}{Lemma}[subsection]
\newtheorem {theorem}{Theorem}[subsection]
\newtheorem {coro}{Corollary}[subsection]
\newtheorem {defi}{Definition}[subsection]
\newtheorem {obs}{Remark}[subsection]
\newtheorem {prop}{Proposition}[subsection]
\newtheorem {exa} {Example} [subsection]

\begin{flushright}
FTUV-00-0613, IFIC-00-22 \\ 
DAMTP-2000-40\\
13th June 2000
\end{flushright}
\vskip 2cm

\begin{center}
{\Large Optimally defined Racah-Casimir operators for $su(n)$} 
\\\vspace{0.1cm}
{\Large  and their eigenvalues for various classes of representations}\\
\vspace{1cm}

\begin{sl}
{\large J.A. de Azc\'arraga$^1$ and A.J. Macfarlane$^2$}\\
\vskip .5cm
$^1$Dpto. de F\'{\i}sica Te\'orica and IFIC, Facultad de Ciencias,\\
46100-Burjassot (Valencia), Spain\\
$^2$Centre for Mathematical Sciences, D.A.M.T.P\\
Wilberforce Road, Cambridge CB3 0WA, UK \\
\vskip 1cm

\vspace{0.5cm}
\end{sl}
\end{center}
\vspace{1cm}
\begin{abstract}
This paper deals with the striking fact that there is an essentially 
canonical path from the $i$-th Lie algebra cohomology cocycle, 
$i=1,2,\dots l$, of a simple compact Lie algebra $\g$ of rank $l$ to 
the definition of its primitive Casimir operators $C^{(i)}$ of
order $m_i$. Thus one obtains a complete set of Racah-Casimir 
operators $C^{(i)}$ for each  $\g$ and nothing else. The paper 
then goes on to develop a general formula for the eigenvalue 
$c^{(i)}$ of each $C^{(i)}$ valid for any representation of  $\g$, 
and thereby to relate $c^{(i)}$ to a suitably defined 
generalised Dynkin index. The form of the formula for $c^{(i)}$
for $su(n)$ is known sufficiently explicitly to make clear some 
interesting and important features. For the purposes of 
illustration, detailed results are displayed for some classes of 
representation of $su(n)$, including all the fundamental ones and 
the adjoint representation.
\end{abstract}

\section{Introduction}

It is well-known 
(see {\it e.g.} \cite{dApb,tensors,cup} for lists of references
and further details) that the $l$ basis elements of the Lie 
algebra cohomology of a simple 
compact Lie algebra $\g$ define, up to a constant, $l$ totally 
antisymmetric tensors. In fact, these may also be understood as 
the coordinates of the different invariant 
$(2m-1)$-forms on the 
manifold of the compact group $G$ associated with $\g$ that, in the 
Chevalley-Eilenberg version of the Lie algebra cohomology \cite{CE},
characterise the ($2m-1$)-cocycles. Given a simple compact $\g$, 
we shall refer to these $l$ tensors 
as the Omega tensors $\Omega^{(2m_s-1)}$ of $\g$. They have orders 
$2m_s-1, \; s=1,2, \dots ,l$, where $m_s$ are the 
orders of the primitive Casimir-Racah operators of $\g$
(see also \cite{dApb,tensors,cup} for lists of references).
For $su(n)$, $m_s \in\{2,3, \dots ,n \}$, and hence 
the Omega tensors are of orders
$3, 5,\dots (2l+1)$. There is an 
essentially canonical path from the Omega tensors 
of a given $\g$ that leads to the set of its $l$ {\it primitive} 
Racah-Casimir operators $C^{(m_s)}$. Following this path ~\cite{tensors}, 
the resulting set of Racah-Casimir operators $C^{(m_s)}$ (represented 
by invariant symmetric tensors $t^{(m_s)}$ of order $m_s$) is optimally 
defined in the sense that it contains one member for each required 
order $m_s$ and nothing else. The procedure allows for the appearance 
of no $C^{(m_s)}$ other the $l$ primitive ones: any formal attempt to 
define  $C^{(m)}$ for, say, $su(n)$ for $m > n$ simply produces 
a vanishing result. Since this paper concentrates on $su(n)$, 
we shall not worry about the refinements that are needed to deal
explicitly with all the invariants of the even orthogonal algebras 
$\g=D_{l}$, where the Pfaffian enters the picture. Nor will the 
subsequent discussion make explicit the qualifications that 
may be needed to cover the exceptional algebras.

The paper proceeds from the definition of a complete 
set of primitive Racah-Casimir operators for $\g$ to a new 
general result for the eigenvalues $c^{(m_s)}(D)$ of
$C^{(m_s)}(D)$ for a generic representation $D$
\be \label{A1} X_i \mapsto D(X_i) \quad \e \noindent
of the Lie algebra
\be \label{A2} {[} X_i \, , \, X_j {]}=if_{ijk} X_k \e \noindent
of $\g$. We have here written $f_{ijk}$ for the structure constants of
$\g=su(n)$. For this algebra, almost all of the technical machinery 
is at hand \cite{tensors, dAMomega} to enable us to display 
explicitly the key features of our general result for $c^{(m)}(D)$. 
Obvious analogues of these results are applicable to all other $\g$.

Our main result states that, 
for any representation $D$,
\be \label{A3}
({\rm dim} \,D)\,c^{(m_s)}(D)=
2^{1-m_s} (gdi)^{(m_s)}(D) \; {\Omega^{(2m_s-1)}}^2 
\e \noindent 
where 
\be 
\label{A4} 
{\Omega^{(2m_s-1)}}^2 \equiv
\Omega_{i_1 \dots i_{2m_s-1}} \Omega_{i_1 \dots i_{2m_s-1}}\quad , 
\e \noindent 
and $(gdi)^{(m_s)}(D)$ is a number dependent 
on the order $m_s$ of the Racah-Casimir, the representation $D$ 
considered and $\g$, or rather in the case of $su(n)$, on $n$. 
For the representation considered,
$(gdi)^{(m_s)}(D)$, $s=1,\dots,l$, is an acronym for 
$s$-th  {\it generalised Dynkin index} for the representation
$D$ considered and its use in (\ref{A3}) is discussed below. 
What is special about $su(n)$ is that 
${\Omega^{(2m_s-1)}}^2$ is known explicitly for all $n$ and 
for all $2\leq m\leq n$ (from now on, we drop the subindex 
$s=1,\dots,l$ in $m_s$). From \cite{dAMomega} we quote
\bea  
{\Omega^{(2m-1)}}^2 & = & \frac{2^{2m-3}}{(2m-2)!} \, n \prod_{r=1}^{m-1} 
(n^2-r^2)\label{A5} \\
& = & \frac{4}{(2m-2)(2m-3)} \; {\Omega^{(2m-3)}}^2 
\quad . \label{A5A}
\ea \noindent 
Eq. (\ref{A5}) exhibits features of (\ref{A3}) which
we believe apply equally well to all other $\g$. Eq. (\ref{A5}) 
shows that ${\Omega^{(2m-1)}}^2 \not= 0$ and hence 
$\Omega^{(2m-1)}$ is non-vanishing only of $m \leq n$. In other 
words the 
primitive ($2m-1$)-cocycle exists only for $m\leq n$
as known from Lie algebra cohomology, and (\ref{A3}) gives a 
null result for $c^{(m)}(D)$ only when
$n< m$. The power of two in (\ref{A3}) 
has been chosen, as far as we know it to be necessary, to ensure 
that, as is customary for an index, $(gdi)^{(m)}(D)$ takes on 
only integral values. In the case of $m=2$ and the familiar Dynkin 
index itself \cite{dynkin}, (\ref{A3}) takes on its standard form 
(see {\it e.g.} \cite{slansky}) 
\be 
\label{A6} 
(gdi)^{(2)}(D)=\frac{2\,{\rm dim} \, D}{n\,{\rm dim} \; \g}
\; c^{(2)}(D) \quad , 
\e 
\noindent
using $f_{ijk}f_{ijk}={\Omega^{(3)}}^2 =n(n^2-1) 
= n.{\rm dim} \,(su(n))$. The factor $n$ in the denominator of (\ref{A6}) 
reflects the fact 
that for uniformity (in $m$) of our definitions of the various
$C^{(m)}$ for $su(n)$, we have defined the quadratic Casimir 
operator of $su(n)$ as
\be 
\label{A7} 
C^{(2)}=n X_i X_i \quad , \quad i=1,\dots,n-1 \quad,
\e \noindent 
see Sec. 2.3. For higher values of $m$ there is less agreement 
as to how the 
Casimir operators $C^{(m)}$, and hence the 
$(gdi)^{(m)}(D)$, should be defined. We have argued that our 
definition of the former is optimal, featuring as it does $t$-tensors 
\cite{tensors} and 
Omega tensors that are in one-to-one correspondence with the 
cohomology cocycles of $su(n)$. This is tantamount to asserting that 
the Omega tensors are the fundamental entities (in fact, $\Omega^3$ 
is given by the structure constants of the 
algebra themselves), 
and our definition of Casimir operators follows from this and 
reflects it too. A good recent account of the role of, and of one 
way of defining, generalised Dynkin indices, which were originally 
introduced in \cite{psw}, is \cite{vRSV}, 
which refers to earlier papers \cite{okubo1} \cite{okpat}. 
The paper \cite{vRSV} also emphasises the role of `orthogonal' 
tensors --in essence our $t$-tensors-- making reference to 
\cite{tensors} in this context, and 
attributing the recognition of the importance of orthogonality 
to the definition of generalised Dynkin indices to \cite{psw}. Ref.
\cite{vRSV} contains extensive tabulations of generalised Dynkin 
indices, as does \cite{psank}. Another useful discussion
of indices is contained in \cite{cvit}, which aims, as we do here, 
at getting results for all $su(n)$ valid for all $n$, using 
procedures --Cvitanovic's bird-track methods-- that are there also applied 
to other $\g$. Our work differs from that of the 
papers just cited in that 
it emphasises the central role of the Omega-tensors, 
and employs a definition of indices that follows 
from this viewpoint. In view of (\ref{A5}) we believe that a 
significant amount of new information is contained in our work. 
We would further wish to advocate that (\ref{A3}) --the formula for
the eigenvualue  $c^{(m)}(D)$ of the Racah-Casimir operator of
oder $m$-- be seen as the result of primary importance. The 
$(gdi)^{(m)}(D)$ are merely numbers,
knowledge of which is required to complete the 
determination of the $c^{(m)}(D)$.
Thus while the number $(gdi)^{(m)}(D)$ in some sense characterises 
the eigenvalue $c^{(m)}(D)$, general $su(n)$ formulas for 
$(gdi)^{(m)}(D)$ do not automatically exhibit the restrictions on $n$ 
necessary for their applicability.
Thus, for the adjoint 
represention $ad$, see below, of $su(n)$, 
\be \label{A8} 
(gdi)^{(4)}(ad)=2n \quad  
\e \noindent
But this applies {\it only} when $n \geq 4$, since (\ref{A3})--(\ref{A5}) 
show that $c^{(4)}(D)$ equals zero for any $D$ for $n=2,3$, as it should do, 
since ${\cal C}^{(4)}$ is absent for these $n$-values. See also comments 
following (\ref{H11}) and (\ref{K9}) below.

The paper turns next to providing some illustrations of 
the results that are contained within (\ref{A3}). We wish to 
deduce the values of $(gdi)^{(m)}(D)$ for various $m$ and 
representations $D$, obtaining results from a single computation 
that are valid for all $n$. For this purpose we consider the 
following classes of representations of $su(n)$. We will use 
the notation ${\cal F}^s$ for $s = 1,2 \dots , l$, for the 
fundamental representations of $su(n)$ of rank $l=n-1$ (writing 
also ${\cal F} \equiv {\cal F}^1$ for the defining representation),
and denote the adjoint representation
of the algebra by $ad$.

$\bullet$ 
{\it The defining representation of $su(n)$, ${\cal F}$} 

${\cal F}=(1,0, \dots ,0)$ and is given by
\be \label{A9} 
X_i \mapsto \fract{1}{2} \lambda_i  \equiv {\cal F}_i \quad ,
\e \noindent
where the $\lambda_i$ are the standard Gell-Mann matrices of $su(n)$ 
\cite{msw}. Using results from \cite{dAMomega}, we can show by a single
calculation that
\be \label{A10}  (gdi)^{(m)}({\cal F})=1 \e \noindent
for all $n$ and for 
all $m \leq n$. This simple general result offers some 
indication of the appropriateness of our definition of $(gdi)^{(m)}$.

$\bullet$ 
The {\it adjoint representation $ad$ of $su(n)$} 

In Dynkin coordinates $ad =(1,0, \dots ,1)$ (see {\it e.g.} 
\cite{slansky}) and it is given by
\be 
\label{A11} 
X_i \mapsto ad (X_i) \equiv ad_i \quad , \quad
(ad_i)_{jk} =-i f_{ijk} \quad . \e \noindent
We show that $(gdi)^{(m)}({\cal F})=0$ for all odd $m$ 
for all $n$, and give results for $m=2,4$ and $6$, on the basis 
of one calculation for each of these three $m$ values.

$\bullet$
{\it The representation $\D$ of $\g = su(n)$}

Let $\D$ denote the representation of $\g$ built \cite{brra} 
\cite{dAM2000} using the set of hermitian Dirac matrices of a  
euclidean space of dimension ${\rm dim}\,\g=r$.
The representation  $\D$ is defined by means of
\be \label{A12} 
X_i \mapsto {\cal D}(X_i)\equiv S_i
=-\fract{1}{4}i f_{ijk} \gamma_j \gamma_k 
\quad ,
\e \noindent 
in terms of Dirac matrices such that
\be \label{A13} \{ \gamma_i \, , \, \gamma_j \} =2\delta_{ij} \quad ;
\e 
\noindent 
hence, $\D$ has dimension $2^{[r/2]}$ 
($[x]$ denotes the integral part of $x$). Unlike 
all the other representations that we 
treat $\D$ is reducible; it describes the direct sum of a number 
of copies of the irreducible representation of $\g$ whose highest 
weight is the principal Weyl vector 
$\delta =\fract{1}{2} \sum_{\alpha> 0} \alpha$, 
{\it i.e.} half the sum of the positive roots of $su(n)$, given by
$\delta=(1,\mathop{\dots}\limits^l ,1)$ in Dynkin coordinates. 
This representation has dimension $2^{(r-l)/2}$.
The actual number of copies of $\delta$ in $\D$ is
$2^{{[}l/2{]}}$. It follows that 
${\rm dim} \, \D=2^{{[}l/2{]}} \; 2^{(r-l)/2}=2^{{[}r/2{]}}$.
We restrict to $\g=su(n)$, for which $l=n-1$ and $r=n^2-1$.
We show that $(gdi)^{(m)}({\cal D})$ is zero for all odd $m$ for 
all $su(n)$, and give explicit results for $m=2$ and $m=4$ only.

$\bullet$ 
{\it The $l$ fundamental representations ${\cal F}^s$ and 
the irrep ${\cal S}_p = (p,0,\dots,0)$}. 

By dealing with the completely symmetrised and 
also the completely antisymmetrised direct products of $p$ 
copies of the defining representation ${\cal F}$, we derive results 
for the representations of $su(n)$ with Dynkin coordinates 
$(p, 0, \dots ,0)$, and for the fundamental representations 
${\cal F}^s, \quad s=1,2,\dots ,n-1$. For the latter, we give
all indices for all ${\cal F}^s$ for $n=3,4,5$ and $6$. 
Bird-track methods \cite{cvit}
are employed here.

The material of this paper is organised as follows. 
Sec. 2 two gives a brief
description of the various families of $su(n)$ tensors, including the
Omega tensors, that are involved in the build up to the 
definition of Casimir operators. To some extent this reviews our 
earlier paper \cite{tensors}, where detailed references to previous 
studies of Casimir operators may be found. In Sec. 3, we derive and 
discuss the key result (\ref{A3}). In Sec. 4, we turn to the 
illustration of (\ref{A3}) for the classes of $su(n)$ representations 
just listed.
 
The interest of this paper in the eigenvalues of Casimir operators has been
in the context of generalised Dynkin indices, because our approach brings 
these in completely naturally. There are however other sources of information
on the subject. There is \cite{okubo} where valuable explicit formulas are 
given for all Casimir operators of all classical algebras and also for 
$g_2$, while \cite{gunkar} addresses the problem for other exceptional 
algebras.

\section{Definitions of tensors and Racah-Casimir operators}

\subsection{The $d$-tensors}

This is a family of symmetric tensors first defined 
by Sudbery \cite{sud}. The 
definition sets out from the standard Gell-Mann totally symmetric 
third order 
tensor $d_{ijk}$ that exists for all $n \geq 3$ and is traceless
$d_{ijk} \delta_{ij}=0$. Higher order tensors in the family 

\be \label{B1} 
d^{(r)}_{(i_1, \dots , i_r)} \quad, 
\e 
\noindent
are defined recursively by symmetrising 
\be \label{B2} d^{(r)}_{i_1, \dots , i_r}=d^{(r-1)}_{i_1, \dots , i_{r-2}j}
d^{(3)}_{j i_{r-1} i_r} \quad , 
\e \noindent 
over all its $i_1,\dots,i_r$ indices. Round 
brackets here denote symmetrisation with unit weight. Thus
\bea 
d^{(3)}_{(ijk)} & = & d_{ijk} \nonumber \\
d^{(4)}_{(ijkl)} & = & \fract{1}{3} (d_{ijp} d_{pkl} +
d_{jkp} d_{pil}+d_{kip} d_{pjl}) \quad . \label{B3} \ea \noindent
Sometimes it is useful to refer to $d^{(2)}_{ij}= \delta_{ij}$ as the rank two
member of the family.

\subsection{The Omega tensors}

Using the mentioned correspondence between $(2m-1)$-cocycles and
$\Omega^{(2m-1)}$ tensors, we have (see, e.g.~\cite{dApb,tensors} 
for the structure of these expressions)

\bea
\Omega^{(3)}_{ijk} & = & f_{ijk} =f_{aij} d^{(2)}_{(ak)} \quad, \label{C1}  \\
\Omega^{(5)}_{ijkpq} & = & f_{a[ij} f^b{}_{kp]} d^{(3)}_{(abq)}\quad, \label{C1A} \\
\Omega^{(7)}_{ijkpqrs} & = & f_{a[ij} f^b{}_{kp} f^c{}_{qr]}  d^{(4)}_{(abcs)}
\quad ,
\label{C2}  \ea \noindent and in general
\be \label{C3} 
\Omega^{(2m-1)}_{i_1 j_i \dots i_{m-1} j_{m-1} k}=
{f^{k_1}}_{[i_1 j_1} \dots {f^{k_{m-1}}}_{i_{m-1} j_{m-1}]} 
d^{(m)}_{(k_1 \dots k_m)} \quad . 
\e \noindent 
Here square brackets indicate unit weight antisymmetrisation over 
all the surrounded indices. The structure
of $\Omega^{(2m-1)}$ above is general for any $\g$; what makes
it specific to $su(n)$ are the orders
($3,5,\dots,(2l+1)$), and of course the fact that the  $f$'s are 
the $su(n)$ ones. One may check explicitly that $\Omega^{(2m-1)}$ tensor
is fully  antisymmetric in all its $(2m-1)$ indices 
$(i_1j_1i_2j_2\dots i_{m-1}j_{m-1}k)$, even if only the first 
$(2m-2)$ indices are antisymmetrised by actual square brackets. The position of
the indices in this paper is without metric signifance. Since $\g$ is compact and its 
generators are hermitian, we may take the Killing metric as the unit matrix, 
and so the raising of indices may just serve to remove them from the sets of 
indices that are subject to antisymmetrisation (or symmetrisation).

A detailed account of the properties of Omega tensors has recently 
been prepared \cite{dAMomega}. 
The extensive compilation of results contained in \cite{dAMomega} 
includes the important formula (\ref{A5}), together with its 
derivation. As noted above, (\ref{A5}) makes clear that
$\Omega^{(2m-1)}$ is absent for $su(n)$ whenever $m > n$.

\subsection{The $t$-tensors}

Following \cite{tensors} we review some properties of this family 
of totally symmetric and totally tracelsss tensors $t^{(m)}$ for $su(n)$. 
The definitions are
\bea
t^{(2)}_{ak} & = & \Omega^{(3)}_{ijk} f_{ija} \quad,\label{D0} \\
t^{(3)}_{abm} & = & \Omega^{(5)}_{ijklm} f_{ija} f_{klb} \quad ,\label{D1} \\
t^{(4)}_{abcq} & = & \Omega^{(7)}_{ijklmpq} f_{ija} f_{klb} f_{mpc}
\quad , \label{D2} \\
t^{(5)}_{abcds} & = & \Omega^{(9)}_{ijklmpqrs} f_{ija} 
f_{klb} f_{mpc} f_{qrd} \quad , 
\label{D2A} 
\ea \noindent 
and in general
\be \label{D3}
t^{(m)}_{k_1 \dots k_m} = \Omega^{(2m-1)}_{i_1 j_i \dots i_{m-1} j_{m-1} k}
f_{i_1 j_1 k_1} \dots f_{i_{m-1} j_{m-1} k_{m-1}}  \quad .
\e \noindent
It follows from definition (\ref{D3}) that the tensor $t^{(m)}$ is fully 
symmetric (\cite{tensors}, lemma 3.2)

   We do not make extensive use of explicit expressions for 
the $t$-tensors, but it is useful
to note results 
from \cite{tensors, dAMomega}
\bea 
 t^{(2)}_{ij} & = & n\delta_{ij} \quad , \label{D3A} \\
 t^{(3)}_{ijk} & = & \frac{1}{3} n^2 d_{ijk} \quad , \label{D4} \\
 t^{(4)}_{ijkl} & = & \frac{1}{15}(n(n^2+1) d^{(4)}_{(ijkl)}
-2(n^2-4) \delta_{(ij}\delta_{kl)}) \quad , \label{D5} \\
 t^{(5)}_{ijklm} & = & \frac{n}{135} ((n(n^2+5) d^{(5)}_{(ijklm)}
-2(3n^2-20) d_{(ijk}\delta_{lm)})   \quad . 
\label{D6} 
\ea 
\noindent
We have adjusted the normalisations in (\ref{D3A})--(\ref{D6}) 
by excluding some powers of two present in (6.12)--(6.14) 
of \cite{tensors}. The 
$t$-tensors are `orthogonal' among themselves, which means that the maximal
contraction of a $t^{(m)}$ with a tensor $t^{(m')}$ of different
order yields zero. This implies, in particular, that $t^{(m)}$ 
is traceless with respect any two indices.
In the simplest, third order case, eq. (\ref{D4}), this is
just the well known property $d_{ikk}=0$. For order four, 
$t^{(4)}$, this means that
\be 
\label{D7} 
t^{(4)}_{ijkl} \delta_{ij}=0 \quad , \quad
t^{(4)}_{ijkl} d_{ijk}=0  \quad .
\e 
\noindent
But, since the trace formulas for $d$-tensors give
\be
\label{otraD} 
d^{(4)}_{(ijkl)} d_{ijm}=\fract{2}{3}
\fract{(n^2-8)}{n} d_{klm} \quad ,
\e \noindent
we learn that the non-trivial result
for the non-maximal contraction
\be \label{D8} t^{(4)}_{ijkl} d_{ijm}=\fract{2}{45} n^2 
(n^2-9) d_{klm} \quad , 
\e \noindent
holds, although it collapses to zero, as it ought,
for $n=3$. Also (\ref{D8}) yields the second part of (\ref{D7}) when one
makes the contraction $k=m$.
For order five, the orthogonality of the $t$-tensors means that
\be 
\label{masD} 
t^{(5)}_{ijklm} {t^{(2)}}_{ij}=0 \quad , \quad
t^{(5)}_{ijklm} {t}^{(3)}_{ijk}=0  \quad , \quad
t^{(5)}_{ijklm} {t^{(4)}}_{ijkl}=0 \quad ,  
\e 
\noindent
and so on. 

One way to see that the  $t$-tensors, like the Omega tensors, 
are absent for $m > n$ is to consider the fully contracted 
square of a generic $t$-tensor
\be 
\label{D9}
{t^{(m)}}^2=t^{(m)}_{k_1 \dots k_m} t^{(m)}_{k_1 \dots k_m} \quad . 
\e \noindent 
This
scalar quantity can be seen to contain the same product of factors 
as is seen in (\ref{A5}). Thus it is a polynomial in $n$ that has factors
which vanish whenever $n<m$. Actually the proof of this claim for the 
$t$-tensors was achieved for $m \leq 5$ in \cite{tensors}, whereas 
(\ref{A5}) is proved in full generality in \cite{dAMomega}, 
relatively speaking rather easily.

The $d$-tensors are less convenient than the $t$-tensors in that 
for $su(n)$ they are well-defined for any order
$m$, but are present as non-primitive tensors for $m>n$. However \cite{sud} 
\cite{tensors} 
\cite{mpf} the unwanted or rather inessential $d$-tensors of
higher orders can always be expressed in terms of the primitive set with
$m \leq n$. For example for $su(3)$ we have
\bea 
d^{(4)}_{(ijkl)} & = & \fract{1}{3} \delta_{(ij} \delta_{kl)} \quad , 
\label{D10} \\
d^{(5)}_{(ijklm)} & = & \fract{1}{3} \delta_{(ij} d_{klm)} 
\quad  \label{D11} \ea \noindent and, for $su(4)$, 
\be \label{D12} 
d^{(5)}_{(ijklm)} = \fract{2}{3} \delta_{(ij} d_{klm)} 
\quad . 
\e \noindent
The above expressions exhibit the non-primitive character of the
symmetric tensors on their left hand sides. It becomes increasingly 
hard to supply such results 
for $su(n)$ at higher $n$. Fortunately this is unnecessary. 
Further, as \cite{dAMomega} shows, the $d$-tensors serve 
perfectly well for the definition of Omega tensors.
If non-primitive $d$-tensors like (\ref{D10})--(\ref{D12})
are employed in the definition of tensors like those of Sec.2.2,
when we know that no Omega tensor 
is allowed (\cite{tensors}, Cor. 3.1), a vanishing result is 
obtained \cite{dAMomega}.
In follows that symmetric invariant tensors differing in 
non-primitive parts lead to the same Omega tensors.

\subsection{The Racah-Casimir operators}

Given the $t$-tensor of 
eq. (\ref{D3}), we define, for $su(n)$, the 
generalised Casimir operator of rank $m$ by means of
\be \label{E1}
C^{(m)}= t^{(m)}_{i_1 i_2 \,  \dots \, i_m} X_{i_i} X_{i_2} \dots
X_{i_m} \quad , \e \noindent
where the $X_i$ are the generators of of the Lie algebra (\ref{A2}) of
$su(n)$. The definition (\ref{E1}) produces each of the primitive 
$su(n)$ Casimir operators of orders $m \in \{ 2,3, \dots , n \} $, 
and nothing else. This is so because the $l$ $su(n)$ 
Lie algebra cohomology cocycles and their associated Omega 
tensors are in one-to-one correspondence 
with the $t$-tensors and hence with the $C^{(m)}$. Had we used 
the $d$-tensors in (\ref{E1}) instead of the $t$-tensors, we would 
always thereby obtain commuting $su(n)$ invariant operators, but of all
orders for all $su(n)$ so that all but
$l$ of them are non-primitive. For low enough $n$, we can derive 
results which show explicitly how some of the non-primitive operators
so obtained can be written in terms of primitive ones. But, 
in the context of the present work, this is not important: 
use of (\ref{E1}) bypasses the problem entirely.

We should point out one consequence of the uniformity in $m$ of the
definitions (\ref{D3A})--(\ref{D6}) of $t$-tensors. Eq. (\ref{D3A}) implies
\be \label{E2}  C^{(2)}=t^{(2)}_{ij} X_i X_j=nX_i X_i \quad , 
\e \noindent
with a possibly unexpected, but harmless, factor $n$. 
For example, for the $su(3)$ representation $(\lambda,\mu)$ 
in Dynkin coordinates, eq. (\ref{E2}) gives the eigenvalue
\be 
\label{E3} 
c^{(2)}(\lambda, \mu)=
(\lambda^2 + \lambda \mu + \mu^2 +3\lambda + 3\mu ) \quad , 
\e \noindent	
and, for the representation $(\lambda,\mu,\nu)$
({\it cf.} \cite{slansky}) of $su(4)$
\be 
\label{E4} 
c^{(2)}(\lambda, \mu, \nu)=\fract{1}{2}
(3\lambda^2 +4\mu^2 +3\nu^2 +4\lambda \mu +2\lambda \nu +4\mu \nu
+12\lambda + 6\mu +12\nu) \quad . 
\e 
\noindent
It may also be worth mentioning the result \cite{klm} for the 
eigenvalue of the cubic Casimir operator of $su(3)$
\be \label{E5}
c^{(3)}(\lambda, \mu)=\fract{1}{6} (\lambda +2\mu +3)(2\lambda +\mu +3)
(\lambda -\mu) \quad. 
\e 
\noindent 
One may use the defining representations of $su(3)$ and $su(4)$ to
check that the normalisations of (\ref{E3})--(\ref{E5}) give agreement
with (\ref{E1}), (\ref{D3A}) and (\ref{D4}).
  
\section{The eigenvalues of the higher Casimir operators}

For $su(n)$ for large enough $n$, 
$n \geq m$, we defined the $m$-th order Casimir 
operator by means of (\ref{E1}), where the $X_i$ 
are the $su(n)$ generators. This yields an invariant operator 
$C^{(m)}(D)$ with eigenvalue $c^{(m)}(D)$ in any representation 
$X_i \mapsto D_i$ so that
\be \label{F1}
C^{(m)}(D)=c^{(m)}(D) \, I_{{\rm dim}\,D} \quad . 
\e \noindent
Then, using (\ref{D3}), we find
\bea 
c^{(m)}(D) \,{\rm dim}\,D & = & {\rm Tr}\, C^{(m)}(D) \nonumber \\
& = & \Omega^{(2m-1)}_{i_1 j_i \dots i_{m-1} j_{m-1} k_m}
f_{i_1 j_1 k_1} \dots f_{i_{m-1} j_{m-1} k_{m-1}}   
{\rm Tr} \, D_{k_1 \dots k_m} \nonumber \\
& = & (-i)^{m-1} \, \Omega^{(2m-1)}_{i_1 j_i \dots i_{m-1} j_{m-1} k_m}
{\rm Tr} \, \left( [D_{i_1},D_{j_1}] \dots [D_{i_{m-1}},D_{j_{m-1}}] 
D_{k_m} \right) \nonumber \\
& = & (-2i)^{m-1} \, \Omega^{(2m-1)}_{i_1 j_i \dots i_{m-1} j_{m-1} k_m}
{\rm Tr} \, D_{[i_1 j_1 \dots i_{m-1} j_{m-1} k_m]} \quad . 
\label{F2} 
\ea \noindent
The first trace in (\ref{F2}), ${\rm Tr} \, D_{k_1 \dots k_m}$
is in practice a unit weight symmetric trace 
${\rm Tr} \, D_{(k_1} \dots D_{k_m)}$
since $t^{(m)}$ is symmetric and $C^{(m)}(D)$ is given by 
(\ref{E1}) with $X=D$. For the last one we have written
\be \label{F3}
D_{[ij \dots s]}= D_{[i} D_j \dots D_{s]} \quad . 
\e 
\noindent
We note here the transfer of the total antisymmetry from the Omega 
tensor to the trace of a product of ($2m-1$) $D$'s. This enables a 
crucial development since, 
as stated, ${\rm Tr}(D_{[i_1j_1\dots i_{m-1}j_{m-1}k_m]})$ must
belong to the $(2m-1)$-cocycle space and, hence, has to be 
proportional to the only primitive, $SU(n)$-invariant,
skewsymmetric, $(2m-1)$-Omega tensor.
 
Hence (see also subsection 3.1) for any representation $D$, we 
may write
\be \label{F4} 
{\rm Tr} \, D_{[i_1 j_1 \dots i_{m-1} j_{m-1} k_m]} = {\left( \fract{1}{4}i
\right)}^{m-1} \, (gdi)^{(m)}(D) \,
\Omega^{(2m-1)}_{i_1 j_i \dots i_{m-1} j_{m-1} k_m} \quad , \e \noindent 
thereby defining a quantity $(gdi)^{(m)}(D)$
which depends on $m$ and $D$ and in general on the Lie algebra $\g$ in 
question. Since $\g=su(n)$ here, $(gdi)^{(m)}(D)$ depends also on $n$.
As noted above $(gdi)$ is an acronym for generalised Dynkin index.
Insertion of (\ref{F4}) into (\ref{F2}) immediately gives rise to 
one of the main results of this paper
\be 
\label{F5}
{\rm dim}\,D \, c^{(m)}(D)=2^{(1-m)} (gdi)^{(m)}(D) {\Omega^{(2m-1)}}^2 
\quad . 
\e \noindent 
The importance of (\ref{F5}) is enhanced for $su(n)$ 
by the availability of the explicit result (\ref{A5}),
valid for all $n$ and for all $m$ relevant to that $n$ value, $m \leq n$.
The relationship of our discussion of $c^{(m)}(D)$ and $(gdi)^{(m)}(D)$
to the work of previous authors is reviewed in the introduction. 
Our presentation conforms fully to this for $m=2$. 
Otherwise our approach differs form that of others in view 
of the primary role in it that is played by the Omega tensors. 
This feature is inherited from \cite{tensors},
but (\ref{A5}) was not known when \cite{tensors} was written.

We believe the analysis described here for the
$A_l$ family of Lie algebras extends to other classical compact
simple algebras, exhibiting similar attractive features, and in some 
fashion to the exceptional algebras. For example, the crucial 
property of `orthogonality' among two $t$-tensors of different
order follows from their general definition in terms of their
respective $\Omega$ cocycle tensors, and does not depend on the 
specific simple $\g$ being considered~\cite{tensors}.
However the corresponding tensor calculus, and the analogue 
of the $su(n)$ $\lambda$-matrix machinery is not yet developed 
sufficiently to produce simple expressions and 
formulas for all simple algebras.

\subsection{Another approach to (\ref{F4})}

We sketch here another means of justifying our use of (\ref{F4}).

We have by now familiar steps
\be \label{F6}
\tr \, D_{[i_1 j_1 \dots i_{m-1} j_{m-1} k_m]}={\left( \fract{1}{2}i
\right)}^{m-1} {f^{k_1}}_{[i_1 j_1} \dots {f^{k_{m-1}}}_{i_{m-1} j_{m-1}]}   
\tr \, D_{k_1 \dots k_m} \quad . \e \noindent
It is legitimate to insert round brackets first to enclose the set 
$k_1 \dots k_{m-1}$ of indices, and then by use of 
the cyclic property of the trace to extend them to enclose 
the full set $k_1 \dots k_m$. Now we may refer to the discussion 
in \cite{tensors} for the construction of a basis for the vector space of  
$su(n)$-invariant symmetric tensors like $\tr \, D_{(k_1 \dots k_m)}$.
The term in the expansion of $\tr \, D_{(k_1 \dots k_m)}$ with respect 
to this basis which involves $d^{(m)}_{(k_1 \dots k_m)}$ is the significant 
one for our argument. Use of it immediately gives rise to a result for 
$\tr \, D_{[i_1 j_1 \dots i_{m-1} j_{m-1} k_m]}$ of the form (\ref{F4}). 
All the other terms of the expansion are made up of symmetrised 
products of lower order $d$-tensors, and give rise to vanishing 
contributions to (\ref{F4}) in view of Jacobi identities, as 
shown in \cite{dAMomega}.

\section{Application to certain classes of representations of $su(n)$}

We consider here several important classes of representations of $su(n)$, 
including the fundamental ones, for which one may provide a 
definition that applies uniformly for all $n$.

\subsection{The fundamental defining representation ${\cal F}$ of $su(n)$}
The representation ${\cal F}$ is defined 
by (\ref{A9})
where the Gell-Mann lambda-matrices \cite{msw} are subject to
\be \label{G2} {\rm Tr} \, \lambda_i =0\ \quad, \quad 
{\rm Tr} \, \lambda_i \lambda_j = 2\delta_{ij} \quad , \quad 
{\lambda_i}^{\dagger}=\lambda_i  
\quad , \e
\be \label{G3}   
\lambda_i\lambda_j=\fract{2}{n} \delta_{ij} + (d+if)_{ijk} \lambda_k
\quad , 
\e \noindent
valid for all $su(n)$, $n\geq 3$.

Using notation like that defined by (\ref{F3}), we quote 
from \cite{dAMomega} the result
for the trace of the fully antisymmetric product of an {\it odd}
number of ($2m-1$) lambda matrices
\be 
\label{G4}
{\rm Tr}\, \lambda_{[i_1 j_1 \dots i_{m-1} j_{m-1} k]}=
2i^{m-1} \Omega_{[i_1 j_1 \dots i_{m-1} j_{m-1} k]} \quad .  
\e 
\noindent
We may now use (\ref{A9}) to substitute ${\cal F}_i$ for 
$\lambda_i$ in (\ref{G4}), and deduce from 
(\ref{F2}), that
\be 
\label{G5}
nc^{(m)}({\cal F}) = 2^{(-m+1)} {\Omega^{(2m-1)}}^2  \e \noindent
so that (\ref{F5}) gives
\be \label{G6} 
(gdi)^{(m)}({\cal F})=1 \quad ,
\e \noindent
which also follows by comparing (\ref{G4}) and (\ref{F4}).

This result applies to all $su(n)$ and 
for any $m \leq n$. Eqs. (\ref{G5}) and the explicit expression (\ref{A5}) 
for ${\Omega^{(2m-1)}}^2$ show that $(gdi)^{(m)}({\cal F})$ is zero 
otherwise. Eq. (\ref{G6}) does not itself provide new 
information (see table 2 of \cite{vRSV}) but (\ref{G5}) 
presents its information in a way that perhaps is.
In this context, it may be worthwhile to display some formulas 
for eigenvalues
in full detail 
\bea 
c^{(2)}({\cal F}) & = & \fract{1}{2} \, (n^2-1) \quad, \label{G7} \\      
c^{(3)}({\cal F}) & = &  \fract{1}{12} \,(n^2-1)(n^2-4) 
         \quad, \label{G8} \\    
c^{(4)}({\cal F}) & = &  \fract{1}{180} \,(n^2-1)(n^2-4)(n^2-9) 
         \quad ,\label{G9} \\
c^{(5)}({\cal F}) & = &  \fract{1}{1680} \,(n^2-1)(n^2-4)(n^2-9)(n^2-16) 
         \quad .         \label{G10} \ea 
The factors that make $c^{(m)}({\cal F})$ vanish when
$m>n$ are visible here: the last factor above is $(n^2-(m-1)^2)$,
and a non-zero result requires $n\geq m$.

\subsection{The adjoint representation $ad$ of $su(n)$}

The adjoint representation $ad$ of the $su(n)$ algebra
is defined by means of
\be \label{H1} 
X_i \mapsto ad_i \quad , \quad (ad_i)_{jk}=-if_{ijk} \quad .\e
\noindent  We do not possess
a general formula for the factor $\mu$ that occurs in
\be \label{H2}
{\rm Tr} \, ad_{[i_1 \dots i_{2m-1}]}= \mu \, \Omega_{[i_1 \dots i_{2m-1}]}
\quad . \e \noindent However it is easy to prove that 
for $m$ {\it odd} the trace in (\ref{H2}) vanishes, so 
that $c^{(m)}(ad )=0$ for all odd $m$. Since 
the matrices of the adjoint representation of any simple Lie 
algebra are antisymmetric, we find {\it e.g.} that
\be 
\label{H3} 
{\rm Tr} \, ad_{[ijkpq]}={\rm Tr} \, (ad_{[ijkpq]})^T=
-{\rm Tr} \, ad_{[qpkji]}=-{\rm Tr} \, ad_{[ijkpq]}=0 \, . 
\e \noindent
For $m$ {\it even} no such conclusion follows: the same steps 
applied to, say, the sevenfold trace do not give zero, because 
now an odd permutation is required at the last step to restore 
the indices to their original order.

It remains to look at the even cases $m=2,4$ and $m=6$, each by a separate 
calculation to get explicit formulas for $c^{(m)}(ad )$ for $su(n)$.
The results are 
\bea 
c^{(2)}(ad ) & = & n^2 \quad ,\label{H4}               \\      
c^{(4)}(ad ) & = &  \fract{2^3}{6!} \, n^2(n^2-4)(n^2-9) 
     \quad, \label{H6} \\
c^{(6)}(ad ) & = &  \fract{2^5}{10!} \, n^2 \prod_{k=2}^{5}(n^2-k^2)
\quad , \label{H7} 
\ea 
\noindent
from which we may conjecture  that, for arbitrary {\it even} $p$,
\be
\label{nueva}
c^{(p)}(ad) = \frac{2^{p-1}}{[2(p-1)]!} \, n^2 \prod_{k=2}^{p-1}(n^2-k^2)
\quad ,
\e \noindent 
whereas $c^{(odd)}(ad)=0$.

The generalised Dynkin indices are then
 
\bea (gdi)^{(2)}(ad ) & = & 2n \label{H8} \quad,   \\
     (gdi)^{(3)}(ad ) & = & 0 \label{H9}  \quad,   \\
     (gdi)^{(4)}(ad ) & = & 2n \label{H10} \quad , \\
     (gdi)^{(5)}(ad ) & = & 0 \label{H10A} \quad,  \\
     (gdi)^{(6)}(ad ) & = & 2n  \quad ,  \label{H11} 
\ea \noindent 
etc. We recall that ${\cal C}^{(6)}$ is absent for $n < 6$ (eq. (\ref{H7}) 
contains explicit factors that reflect this), and hence note that 
(\ref{H11}) really only applies when $n \geq 6$. See also  
remarks that follow (\ref{K9}). Results (\ref{H8}) and (\ref{H10}) 
agree with results in \cite{cvit}.

The proof of (\ref{H4}) is easy. To obtain (\ref{H6})
we use
\be \label{H12}
{\rm Tr} \, ad_{[ijklpqr]}={\left(\fract{1}{2}i\right)}^3 
{f^a}_{[ij} {f^b}_{kl} {f^c}_{pq]} 
{\rm Tr} (ad_r \, ad_{(abc)}) ={\left(\fract{1}{4}i\right)}^3 
\, 2n \, \Omega_{ijklpqr} \quad . 
\e \noindent
To perform the last step, a result from \cite{tensors} is employed
\be \label{H12A}
{\rm Tr} \, \left( ad_r \, ad_{(abc)} \right)
=\fract{n}{4} d^{(4)}_{(abcr)} +2\delta_{(ab} \delta_{cr)}  \quad . \e
\noindent
In fact the second term of (\ref{H12A}) does not contribute to 
(\ref{H12}) because it is non-primitive.

The proof of (\ref{H7}) similarly requires the formula
\be \label{H13} 
{\rm Tr} \, ad_{(abcder)}= \fract{n}{16} \, d^{(6)}_{(abcder)}+ \dots
\quad , 
\e \noindent where the dots denote terms which do not contribute to
(\ref{H7}). One obtains this result by a method similar 
to that sketched in \cite{tensors} to derive (A21) there. 
This requires a preliminary result 
\be \label{H14}
\tr \, ad _{(abcd} D_{e)} = \fract{n}{8} \, d^{(5)}_{(abcde)}+
\delta_{(ab}d_{cde)}
 \quad , 
\e \noindent 
where $(D_i)_{jk} =d_{ijk}$. The deduction of each of the last 
two results entails a considerable amount of effort, making 
liberal use of identities found in the appendix to \cite{tensors}.

\subsection{The reducible representation ${\cal D}$}

The representation ${\cal D}$ of $\g=su(n)$, 
of dimension $2^{[\rm{dim}\,\g/2]}=2^{[(n^2-1)/2]}$, 
has been described in the introduction. However in this 
case again, we lack an explicit analogue of (\ref{G4}). 
Again too the odd order Casimir operators have zero eigenvalues. 
To show this is true, we note there exists a matrix $C$ such that
\be \label{I1} 
C \gamma_i C^{-1}=\pm {\gamma_i}^T \quad , 
\e \noindent
with the sign depending on ${\rm dim} \,\g$. Hence in general
$(S_i\equiv {\cal D}(X_i))$
\be \label{I2} 
S_i=-\fract{1}{4}i f_{ijk} \gamma_j \gamma_k \quad ,\quad
[S_i,S_j]=f_{ijk}S_k \quad, 
\e \noindent
obeys
\be \label{I3}
{S_i}^T= -CS_iC^{-1} \quad . \e \noindent
This is sufficient to allow steps like those used in (\ref{H2}) to 
complete the demonstration, since the matrix $C$ is invisible 
within the trace.
To get non-vanishing results we look at $m$ even, this time 
confining ourselves to the cases $m=2$ and $m=4$.
We have 
\bea 
 c^{(2)}({\cal D}) & = &   \fract{n}{8} {\Omega^{(3)}}^2    \label{I4} \\ 
 c^{(3)}({\cal D}) & = &  0                       \label{I5} \\    
 c^{(4)}({\cal D}) & = &  -\fract{n}{64}  {\Omega^{(7)}}^2 
         \quad , \label{I6} \ea \noindent
and hence
\bea 
     (gdi)^{(2)}({\cal D}) & = & \fract{n}{4} ({\rm dim}\,{\cal D}) 
\label{I7} \\
     (gdi)^{(3)}({\cal D}) & = & 0 \label{I8} \\
     (gdi)^{(4)}({\cal D}) & = &  -\fract{n}{8} ({\rm dim}\,{\cal D}) \quad . 
          \label{I9} \ea \noindent
We have already noted that ${\cal D}$ is a direct sum of 
$2^{{[}(n-1)/2{]}}$ copies of the irreducible representation 
$\delta=(1, \dots ,1)$ of $su(n)$ \cite{dAM2000},
and that ${\rm dim} \,{\cal D} =2^{{[}(n^2-1)/2{]}}$. 
It follows that the indices given by (\ref{I7}) and  (\ref{I9}) 
are in all cases integers.

We remark also that the results (\ref{I4})--(\ref{I6}) apply 
also to the representation $\delta$ of $su(n)$
since ${\cal D}$ is a direct sum of copies of $\delta$.

For $su(3)$, for which $\delta$ coincides with the adjoint
representation, ${\cal D}$ comprises two copies of $ad$, 
$c^{(2)}(\D)=c^{(2)}(\delta)=c^{(2)}(ad)=9$ (eq. (\ref{H4})), 
but
\be \label{I10} 
(gdi)^{(2)}(\D) = 2 (gdi)^{(2)}(ad) \quad , 
\e 
\noindent
as is to be expected since
by eq. (\ref{F4}) the dimension of the
representation enters into the definition of the Dynkin index.
 
The easier of the proofs known to us for (\ref{I6}) follows the 
same lines as the proof of (\ref{H4}). It therefore requires the result
\be \label{I11} 
{\rm Tr} \, \left( S_d S_{(abc)} \right)
=-\fract{n}{64} d^{(4)}_{(abcd)} ({\rm dim} \,{\cal D})+
\fract{3n^2-8}{64} \delta_{(ab} \delta_{cd)} ({\rm dim} \,{\cal D}) 
\quad , \e
\noindent 
which is proved in much the same way as (A.11) in \cite{tensors} is
proved. Some non-trivial work on traces of gamma matrices is 
involved. Also, as for (\ref{H12A}), the second term of (\ref{I11}) 
does not contribute to the derivation of (\ref{I9}). The minus sign 
in (\ref{I9}) may be noted. 
It is not exceptional: the tables of \cite{vRSV} have plenty 
of negative entries.

\subsection{The representations ${\cal S}_p$ of highest weight 
$(p,0, \dots ,0)$}

The representations ${\cal S}_p$ carried by totally symmetric 
$su(n)$ tensors of rank $p$, the defining representation ${\cal F}$ 
being the case with $p=1$, {\it i.e.} ${\cal S}_1={\cal F}={\cal F}^1$.
We can extend the results obtained for ${\cal F}= (1,0, \dots ,0)$ 
easily to ${\cal S}_2 $ for which we define matrices
\be \label{J1}
{(M_i)}_{\, a_1 a_2, b_1 b_2}= {\delta_{(a_1}}^{(b_1} 
{{\lambda_{i\,}}_{a_2)}}^{b_2)}
\quad . \e \noindent
It is easy to check that (\ref{J1}) satisfies (\ref{A2}). As 
previous sections indicate, what one needs for the calculation of 
generalised Dynkin indices in our approach is the evaluation of traces 
\be 
\label{J2} 
{\rm Tr} \, D_{(a} D_b \dots D_{s)} \quad . 
\e \noindent
It is easy to use (\ref{J1}) to compute

\bea 
{\rm Tr} \, M_i M_j & = & \fract{1}{2}(n+2) \delta_{ij} \label{J3} \\
{\rm Tr} \, M_{(i} M_j M_{k)}  & = & \fract{1}{4}(n+4) d_{ijk} \label{J4} \\
{\rm Tr} \, M_{(i} M_j M_k M_{l)} & = & \fract{1}{8}(n+8) d^{(4)}_{(ijkl)}
+ \dots \quad ,
 \label{J5} \ea \noindent where the dots indicate lower order 
terms known but known also, because of Jacobi identities, not to contribute to 
the calculation of the eigenvalues $c^{(4)}({\cal S}_2)$. Thus 
we find that all results agree with
\be \label{J6}
(gdi)^{(m)}({\cal S}_2) = n+2^{m-1} \quad . \e \noindent

To proceed further it is advisable to use heavier duty methods. Bird-track 
methods allow us to subsume the calculations just done into the treatment of 
the general $p$ case, by dealing with the totally symmetrised $p$-fold direct 
products of defining representations.

Our results include the following
\bea
(gdi)^{(m)}({\cal S}_p) & = & \frac{(n+p)!}{(p-1)!(n+1)!} \quad , \quad
m=2 \label{J7} \quad ;\\
& = & \frac{(n+p)!}{(p-1)!(n+2)!}(n+2p)  \quad , \quad
m=3 \label{J8} \quad ;\\
& = & \frac{(n+p)!}{(p-1)!(n+3)!}(n^2-n+6pn+6p^2) \quad , \quad
m=4 \quad .\label{J9} 
\ea 
\noindent
The result (\ref{J6}) for ${\cal S}_2$ of course conforms 
to these results. To derive these expressions, we employ 
results for totally symmetrised traces that appear in \cite{cvit} 
as eqs. (12.69)--(12.71). We note also from \cite{cvit} the 
diagram (12.64) used to define the matrices of $(p,0, \dots ,0)$, 
and the essential results (5.19) and (5.23) given in the valuable 
chapter in \cite{cvit} on permutations. 

\subsection{The $l$ fundamental representations ${\cal F}^s$ of $su(n)$}

The representation ${\cal F}^2=(0,1,0, \dots ,0)$ of $su(n)$ is the 
antisymmetric part of the direct product ${\cal F} \otimes {\cal F}$, where
${\cal F}= {\cal F}^1$ is the defining representation of $su(n)$. Thus we
define for ${\cal F}^2$ the matrices
\be \label{K1}
(N_i)_{\, a_1 a_2, b_1 b_2}= {\delta_{[a_1}}^{[b_1} 
{{\lambda_{i\,}}_{a_2]}}^{b_2]}
\quad . \e \noindent
This differs form (\ref{J1}) only in that round symmetrisation brackets 
are replaced by square antisymmetrisation brackets.
We find the following results
\bea
{\rm Tr} \, N_i N_j & = & \fract{1}{2}(n-2) \delta_{ij} \label{K3} \\
{\rm Tr} \, N_{(i} N_j N_{k)}  & = & \fract{1}{4}(n-4) d_{ijk} \label{K4} \\
{\rm Tr} \, N_{(i} N_j N_k N_{l)} & = & \fract{1}{8}(n-8) d^{(4)}_{(ijkl)}
+ \dots \quad .
\label{K4A} \ea \noindent 
Again all these results agree with the general statement
\be \label{K5}
(gdi)^{(m)}({\cal F}^2) = n-2^{m-1} \quad . \e \noindent
 
It is of clear interest to proceed further down the antisymmetrisation
path. For $su(n)$, the totally antisymmetrised parts of the 
$s$-fold products of defining representations correspond 
to the fundamental representations ${\cal F}^s$ of $su(n)$ for 
$s=1,2, \dots l=(n-1)$, {\it i.e.} ${\cal F}^s$ has a one in the 
$s$-th place of its Dynkin coordinate description and zeros 
elsewhere: its highest weight is the $s$-th fundamental dominant 
weight. 

Using bird-track methods, we find
\bea
(gdi)^{(m)}({\cal F}^s) & = & \frac{(n-2)!}{(s-1)!(n-s-1)!} \quad , \quad
m=2 \label{K6} \quad ;\\
& = & \frac{(n-3)!}{(s-1)!(n-s-1)!}(n-2s)  \quad , \quad
m=3 \label{K7} \quad ;\\
& = & \frac{(n-4)!}{(s-1)!(n-s-1)!}(n^2+n-6sn+6s^2) \quad , \quad
m=4 \quad .\label{K8} \ea \noindent
The result (\ref{K5}) for $s=2$ conforms to these results; for $s$=1 we get 
$(gdi)^{(m)}({\cal F}^1)$=1, which also follows from
(\ref{J7})--(\ref{J9}) for $p$=1 as it should.  Results 
analogous to (\ref{K6})--(\ref{K8}) for 
$m=5$ and $m=6$ have also been computed. For $s=2$ they 
each agree with (\ref{K5}). When $s=3$ we have
\be \label{K9}
(gdi)^{(5)}({\cal F}_3)=\fract{1}{2} (n-6)(n-27) \quad.
\e 
\noindent 
Since $su(n)$ has a fifth order Casimir operator only for $n \geq 5$,
(\ref{K9}) applies only for such $n$. It gives $11$ for $n=5$ and vanishes
for $n=6$, but is non-zero for all larger n except $n=27$. In other words
$c^{(5)}({\cal F}_3)$ vanishes when $n=6$ in virtue of its $(gdi)$ factor 
rather than its $\Omega$ factor.

To obtain these results we have followed methods for the antisymmetric 
case analogous to those of the previous section for the symmetric case.
Permutation lemmas (5.19) and (5.23) of \cite{cvit} expedite the 
work.

The results of this section permit the evaluation of all the indices of all the
fundamental representations of $su(n)$ for all $n \leq 6$. These are presented
in tables in Sec. 5.

The results of (\ref{J7})--(\ref{J9})  are very closely related to 
results to be found in chapter 16 of \cite{cvit}. Although no such 
statement holds for (\ref{K6})--(\ref{K8}), all the tools needed to 
derive them were found in \cite{cvit} polished and ready for use.

\section{Tables of indices for $su(n)$ for $n \leq 6$}

The $(gdi)$ indices presented in the tables that follow have 
been deduced from (\ref{K5})--(\ref{K9}) one easy check is 
available. If $X_i \mapsto D_i$ defines the representation $D$ 
of $su(n)$, then
\be \label{LO} 
X_i \mapsto {\bar D}_i=-{D_i}^T \quad
\e
defines the representation ${\bar D}$ of $su(n)$. It then follows that
\be 
\label{L1}
c^{(m)}(D)=\pm c^{(m)}({\bar D}) \quad ,
\e 
\noindent
where the plus applies to even $m$ and the minus to $m$ odd. 
The data in the tables below conforms to this. Further
some entries for $su(4)$ and $su(6)$ agree with 
the consequence of (\ref{L1}) that odd Casimir operators have zero 
eigenvalues for self-conjugate representations.

\vspace{1cm}
\begin{center}
\begin{tabular}{|c|c|c|}  \hline
    \multicolumn{3}{|c|}{Generalised Dynkin indices of $su(3)$} \\ \hline
    $su(3)$ & $s=1$ or (1,0) & $s=2$ or (0,1) \\ \hline \hline
    $m=2$   &  1             & 1       \\ \hline
    $m=3$   &  1             & $-1$      \\ \hline
\end{tabular} \\
\vspace{1cm}
\begin{tabular}{|c|c|c|c|} \hline
    \multicolumn{4}{|c|}{Generalised Dynkin indices of $su(4)$} \\ \hline
   $su(4)$ & $s=1$ or (1,0,0) & $s=2$ or (0,1,0) &$s=3$ or (0,0,1) \\ 
             \hline \hline
   $m=2$   &  1             &  2                  & 1       \\ \hline
   $m=3$   &  1             &  0                  & $-1$    \\ \hline
   $m=4$   &  1             &  $-4$               & 1       \\ \hline
\end{tabular} \\
\vspace{1cm}
\begin{tabular}{|c|c|c|c|c|} \hline
   \multicolumn{5}{|c|}{Generalised Dynkin indices of $su(5)$} \\ \hline
   $su(5)$ & $s=1$ or (1,0,0,0) & $s=2$ or (0,1,0,0) &$s=3$ or (0,0,1,0) 
              &$s=4$ or (0,0,0,1) \\ \hline \hline
   $m=2$   &  1             &  3     & 3          & 1       \\ \hline
   $m=3$   &  1             &  1     & $-1$       & $-1$    \\ \hline
   $m=4$   &  1             &  $-3$  & $-3$       & 1       \\ \hline
   $m=5$   &  1             &  $-11$ & 11         & $-1$    \\ \hline
\end{tabular} \\
\vspace{1cm}
\begin{tabular}{|c|c|c|c|c|c|} \hline
   \multicolumn{6}{|c|}{Generalised Dynkin indices of $su(6)$} \\ \hline
   $su(6)$ & (1,0,0,0,0) & (0,1,0,0,0) &(0,0,1,0,0) 
              &(0,0,0,1,0) &(0,0,0,0,1)\\ \hline \hline
   $m=2$   &  1     &  4      &  6     & 4          & 1       \\ \hline
   $m=3$   &  1     &  2      &  0     & $-2$       & $-1$    \\ \hline
   $m=4$   &  1     &  $-2$   &  $-6$  & $-2$          & 1       \\ \hline
   $m=5$   &  1     &  $-10$  &  0     & 10         & $-1$    \\ \hline
   $m=6$   &  1     &  $-26$  &  66     & $-26$      & 1    \\ \hline
\end{tabular} \\

\end{center}

\vskip 1.5cm
{\bf Acknowledgements}. This work was partly supported by the
DGICYT, Spain ($\#$PB 96-0756) and PPARC, UK. AJM thanks the Dpto. de 
F\'{\i}sica T\'eorica, University of Valencia, for hospitality during two 
recent visits when research on the present paper was performed.

\end{document}